\documentstyle[12pt,fleqn]{article}
\begin{document}

\begin{center}
{\large{\bf Isoscalar Roper excitation in the $p p \rightarrow p p \pi^0$
reaction close to threshold}}
\end{center}

\vspace{1cm}

\begin{center}
{\large E. Hern\'andez$^a$ and E. Oset$^b$}
\end{center}

\vspace{0.5cm}

$^a$ {\small{\it Grupo de F\'{\i}sica Nuclear, Facultad de Ciencias
37008 Salamanca, Spain}}

$^b$ {\small{\it Departamento de F\'{\i}sica Te\'orica and IFIC,
Centro Mixto Universidad de Valencia-CSIC,
46100 Burjassot (Valencia), Spain}}
\vspace{3cm}

\centerline{ABSTRACT}
\begin{center}
\begin{minipage}[t]{12.7cm}
%{\footnotesize 
A new mechanism for the $p p \rightarrow p p \pi^0$
reaction close to threshold is suggested coming from the isoscalar
excitation of the Roper and its decay into $N (\pi\pi)_{s-wave}$, with one of
the $\pi^0$ emitted and the other one reabsorbed on the second nucleon. The
 mechanism can lead to 
important interference with other mechanisms and, together with experiment,
serves to  exclude large ranges of the $2\pi$ $N^*$ decay parameters
allowed by the $N^*$ partial decay widths.
%}
\end{minipage}
\end{center}

\newpage

\section{Introduction}
The large discrepancies between early calculations of the 
$p p \rightarrow p p \pi^0$ cross section close to threshold, based on the
one body mechanism and the rescattering term \cite{1,2,3,4}, and the
experimental data \cite{5,6} have stimulated further work looking for a
solution to the puzzle. Short range contributions associated to the
isoscalar excitation of negative energy components on the nucleons were
suggested as a possible explanation of the puzzle \cite{7,8} and similar
ideas using exchange currents with heavy mesons have also been discussed
\cite{9}.

It was soon realized \cite{10} that because the rescattering process 
involves the isoscalar $\pi N$ amplitude around threshold, and this
amplitude is abnormally small on shell, off shell effects should be
relevant since the $\pi N$ amplitude appears there half off shell. 
Quantitative evaluations were done in \cite{11} using two different off
shell extrapolations in order to estimate uncertainties and it was shown
that the use of the off shell extrapolations enhanced appreciably the
cross section and could by itself explain the data. Further work along
these lines was done in \cite{12} improving on the small one body mechanism
and using a different off shell extrapolation obtained from one version of
the Bonn meson exchange model for the $\pi N$ interaction \cite{13}.

The works of \cite{11} and \cite{12} share many things in common, with
quantitative differences mostly due to the different off shell extrapolations
used. In both cases a substantial increase of the cross section is found
due to off shell effects (smaller in ref. \cite{12}), together with a
constructive interference between the one body and the rescattering terms.

The realization that QCD at low energies can be effectively taken into
account by means of effective chiral Lagrangians \cite{14} has led to the
developments of Chiral Perturbation Theory ($\chi$PT) \cite{15,16,17,18},
providing, in principle, and ideal tool to tackle the problem of the off
shell extrapolation in the $p p \rightarrow p p \pi^0$ process. This has
led to some work along these lines \cite{9,19,20} with a main common
feature, with respect to \cite{11,12}, which is the negative interference
between the one body and rescattering terms, opposite to the findings of
\cite{11,12}. Another difference is the small cross sections obtained along
these lines. This approach has been further revised in \cite{21} where the
authors note that several approximations done in the coordinate space 
treatment of former chiral approaches induced large uncertainties. The
improved work of \cite{21} in momentum space produces a much larger 
rescattering term and consequently larger cross sections. Yet, the 
interference between the one body and rescattering terms is negative as in
former approaches.

Further clarifications on the chiral approach appear in the recent paper
\cite{22} which concludes that present $\chi$PT calculations are not
yet at the level of providing quantitative results for the rescattering
term. The large size of the momentum involved in the half off shell
$\pi N$ amplitude requires the evaluation of higher loops, and their
corresponding counterterms. Actually, an accurate evaluation of this
amplitude might as well require the use of non perturbative unitary
techniques with coupled channels, as done in \cite{23,24,25}. A very
accurate description of $K^- p$ scattering going to $K^- p, \bar{K}^0 n,
\Sigma^+ \pi^-, \Sigma^- \pi^+$ and $\Lambda \pi^0$, together with the
dynamical generation of the $\Lambda (1405)$ resonance below $K^- p$
threshold was obtained in these works. One of the findings of \cite{25}
was the relevance of including the $\eta \Lambda$ and $\eta \Sigma^0$
channels in the approach, even if they are not open at low $K^-$ energies,
with some cross sections increased by a factor three due to the inclusion
of these channels. This hints that the inclusion of coupled channels in the
$\pi N$ interaction might be relevant even at pion threshold. Another result
in \cite{25} was the realization that $SU (3)$ symmetry, in the limit of 
equal masses, is broken unless all coupled channels from the octets of 
$1/2^+$ baryons and $0^-$ mesons are included in the coupled channel approach.

Further work along the lines of $\chi$PT is carried out in \cite{26}. In this 
case pionic loops, including two pion exchange diagrams that might simulate
$\sigma$ exchange decaying into two $\pi^0$, one of which is emitted and the
other one reabsorbed into the second nucleon, are included. An excellent
agreement with experiment is claimed, even when the one body term is excluded.
The same occurs in a OBE model by the same authors which explicitly accounts
for the mechanism described above \cite{27}, which leads the authors to claim
that this is the basic mechanism describing the process, irrespective of the
formalism chosen. Other OBE models, not including that latter mechanism also
claim to reproduce the data for $N N \rightarrow N N \pi$ in different isospin
channels \cite{28}.

Undoubtedly much progress is being made, but the main conclusion might be that
the process is more complicated than originally thought and that much work
remains to be done.

The present work calls the attention on new mechanisms, not yet explored, and
that could be relevant for the $p p \rightarrow p p \pi^0$
reaction, when considered in connection with the rescattering term, due
to interference. The mechanism is related to Roper excitation
and its decay into $N (\pi \pi)^{I = 0}_{s = \hbox{wave}}$. This mechanism is
present in most $2 \pi$ production processes around threshold, 
$\pi N \rightarrow \pi \pi N$ \cite{29,30}, $\gamma N \rightarrow \pi \pi N$
\cite{31} and $N N \rightarrow N N \pi \pi$ \cite{32}. In this latter process
this mechanism is by far the dominant one around threshold in 
$p p \rightarrow p p \pi^+ \pi^-$, $p p \pi^0 \pi^0$, where the two pions can
be in $I = 0$ and, within uncertainties, the agreement with data is acceptable.
This gives us some confidence about the size of the mechanism evaluated here
which corresponds to the dominant one for $p p \rightarrow p p \pi^0 \pi^0$
in which one $\pi^0$ is emitted and the other one reabsorbed on the second
nucleon.

\section{Isoscalar Roper excitation in the $N N \rightarrow N N^*$ reaction}

The clean experimental signal for $N^* (1440)$ excitation in $(\alpha, \alpha')$
collisions on a proton target \cite{33} provided evidence of a strong 
isoscalar excitation of the Roper in N N collisions. The experiment was
analyzed in \cite{34} by means of a model which included $\Delta$ excitation
in the projectile (fig. 1a) together with Roper excitation on the target
(fig. 1b), including the interference of both terms (for the part of
$N^* \rightarrow N \pi$). For the isoscalar excitation of the Roper in diagram
1b an empirical amplitude was constructed assuming an effective $``\sigma"$
exchange (although in a more microscopical picture it would be a combination
of $\sigma$ and $\omega$ exchange). This effective $\sigma$ was assumed to
have the same coupling to NN as in the Bonn model \cite{35} while the coupling
to $N N^*$ was fitted to the data. The couplings used were

\begin{equation}
\frac{g^2_{\sigma N N}}{4 \pi} = 5.69 \; ; \;
\frac{g^2_{\sigma N N^{*}}}{4 \pi} = 1.33
\end{equation}

\noindent
and a monopole form factor with $\Lambda_{\sigma} = 1.7 \, GeV$ together with
$m_{\sigma} = 550 \, MeV$, as in \cite{35} were used. With this input, which
contains $g_{\sigma N N^*}$ as the only parameter, a good reproduction of the
data of \cite{33} was obtained.

In the $N N \rightarrow N N \pi \pi$ reaction studied in \cite{32} the same
input for the isoscalar Roper excitation was used and the diagrams of fig. 2,
together with the corresponding ones with $N^*$ excitation on the first
nucleon, plus 13 other mechanisms, including $\Delta$ excitation and chiral
terms, were used. Contrary to the case of the $(\alpha, \alpha')$ reaction where only
the isoscalar exchange is allowed, here we can also exchange an $I=1$
object, but is was shown in \cite{32} that the strength of the isoscalar
exchange was much larger than the corresponding one with $I = 1$, so here
only the isoscalar excitation is considered. The results of \cite{32} showed
that in the $p p \rightarrow p p \pi^+ \pi^-$ and 
$p p \rightarrow p p \pi^0 \pi^0$ reactions the mechanisms of fig. 2 with
$N^* \rightarrow N (\pi \pi)^{I = 0}_{s-wave}$ dominated the cross
sections close to threshold, where the other mechanisms either vanished or
became very small.

With all this previous work described, there is then a clear mechanism which
could be relevant for $p p \rightarrow p p \pi^0$ close to threshold and 
this is the one depicted in fig. 3, which  corresponds to the same mechanism
of fig. 2 for $2 \pi^0$ production, where one of the pions is reabsorbed on
the first nucleon, giving rise to the box diagrams of the figure. This is the
mechanism which we evaluate in the next section.

\section{Box diagram with isoscalar $N^*$ excitation}

For the evaluation of the box diagrams of fig. 3 we need 
the following Lagrangians:
\begin{eqnarray*}
{\cal L}_{\sigma p p}(x)=g_{\sigma NN}\ \bar{\Psi}_p (x)\Psi_p (x)\ \sigma (x)
\end{eqnarray*}
\begin{eqnarray*}
{\cal L}_{\pi^0 p p }(x)=\frac{f_{\pi NN}}{m_{\pi}}\ \bar{\Psi}_p (x) 
\gamma^{\mu} \gamma_5\Psi_p (x)\ \partial_{\mu}\pi^0 (x)
\end{eqnarray*}
\begin{eqnarray*}
{\cal L}_{\sigma p N^*}(x)=g_{\sigma N N^*}\ \bar{\Psi}_{N^*} (x) 
\Psi_p (x)\ \sigma (x) + h.c.
\end{eqnarray*}
\begin{eqnarray}
{\cal L}_{\pi^0 \pi^0 p N^*}(x) &=& g_{1 \pi \pi N N^*}\ 
\frac{m_{\pi}^2}{f_{\pi}^2}\ \bar{\Psi}_{N^*} (x) 
\Psi_p (x)\ \pi^0 (x) \pi^0 (x)  \nonumber \\
& & +\ g_{2 \pi \pi N N^*}\ \frac{1}{f_{\pi}^2}\ \bar{\Psi}_{N^*} (x) 
\Psi_p (x)\ \partial^0 \pi^0 (x)\ \partial^0 \pi^0 (x) + h.c.
\end{eqnarray}

The lagrangian ${\cal L}_{\pi^0 \pi^0 p N^*}(x)$ , with the
second piece in its Lorentz
covariant form, was first used in ref. \cite{30} to evaluate the decay
$N^*(1440) \rightarrow N (\pi \pi)_S$. In \cite{30}
the couplings $g_{1 \pi \pi N N^*}$ and $g_{2 \pi \pi N N^*}$ are called
respectively $-c_1^*$ and $-c_2^*$.
 In this latter lagrangian we have set the energy of the Roper equal to its mass
 with respect to the formal one written in \cite{30}. This is a good
 approximation in the present case.
 In that lagrangian $f_{\pi}$ 
stands
for the pion decay constant $f_{\pi}=92.4 \ MeV$.

The couplings  $ g_{1 \pi \pi N N^*}$ and $g_{2 \pi \pi N N^*}$ are not
fully known. The main constraint to their values 
comes from the study of the decay  $N^*(1440) \rightarrow N (\pi \pi)_S$.
In ref. \cite{32} it is found that:
\begin{eqnarray}
\Gamma_{N (\pi \pi)_S} = \alpha\ 
g_{1 \pi \pi N N^*}^{\ 2}
+\ \beta g_{2 \pi \pi N N^*}^{\ 2} 
+\ \gamma\ g_{1 \pi \pi N N^*}\ 
g_{2 \pi \pi N N^*}
\end{eqnarray}
where  $\alpha=0.497\ 10^{-3}\ GeV^3$, 
$\beta=3.66\ 10^{-3}\ GeV^3$ and $\gamma=2.69\ 10^{-3}\ GeV^3$. For
$\Gamma_{N (\pi \pi)_S}$ they use a branching ratio of $7.5\%$ and a total
width of $350\ MeV$. The above ellipse  is not able by itself to fix both 
parameters and in fact, as seen in fig. 4, it spans over a large range of 
values.
Further constraints were obtained in ref.\cite{32} from an analysis of the
$\pi^- p \rightarrow \pi^+ \pi^- n$ reaction data. Within the model used
 the experiment  seemed to favour intermediate values 
in the ellipse. 
%That intermediate 
%region is the one we will explore in the present calculation. 
We also point out 
 here that in ref. \cite{30} the signs of both $g_{1 \pi \pi N N^*}$
and $g_{2 \pi \pi N N^*}$ are taken to be the same as the ones for the
corresponding couplings in the $NN \pi \pi$ vertices. 
In this paper we will leave 
open the possibility for a different  signs assignment.
               
The net contribution of the four diagrams to the invariant amplitude 
is given by
\begin{eqnarray}
{\cal M}&=&-2\ i\ g_{\sigma NN}\ g_{\sigma N N^*}\ \frac{f_{\pi NN}}{m_{\pi}} 
\ \int \frac{d^4q}{(2 \pi)^4}\ \left( \ g_{1 \pi \pi N N^*}\ m_{\pi}^2-
g_{2 \pi \pi N N^*}\ q^0 p_{\pi}^0 \right)\ \frac{1}{f_{\pi}^2} \nonumber \\
& & \times \ D_{\pi}(q)\ D_{\sigma} (p_3-p_1-q)\ \left( F_{\pi}(q)\right)^2 
\left( F_{\sigma}(p_3-p_1-q)\right)^2  \nonumber \\
& & \times \ \bar{u}_{s3}(\vec{p}_3)\left(\gamma^{\mu} 
\gamma_5 S_p(p_3-q)+
S_p(p_1+q) \gamma^{\mu} \gamma_5\right) u_{s1}(\vec{p}_1)\nonumber \\
& & \times \left\{ \bar{u}_{s4}(\vec{p}_4) 
S_{N^*}(p_4+p_{\pi}+q) u_{s2}(\vec{p}_2) 
+ \bar{u}_{s4}(\vec{p}_4) 
S_{N^*}(p_2-p_{\pi}-q) u_{s2}(\vec{p}_2) \right\}\nonumber \\[.2cm]
& & + \ (\ exchange \ diagrams\ ) 
\end{eqnarray}
\noindent
where we have included monopole form factors $F_{\sigma}$ and $F_{\pi}$ 
for each of the sigma and pion vertices. For the latter we use  
$\Lambda_{\pi}=1.25 \, GeV$. 
For the nucleon and Roper propagators we will take the positive energy 
part alone through the decomposition:
\begin{eqnarray}
S(p)=\frac{1}{2 E(\vec{p})}
\frac{E(\vec{p}) \gamma^0 - \vec{p}\ \vec{\gamma} +m}{p^0-E(\vec{p})+i \epsilon}
+\frac{1}{2 E(\vec{p})}
\frac{E(\vec{p}) \gamma^0 + \vec{p}\ \vec{\gamma} +m}{p^0+E(\vec{p})+i \epsilon}
\end{eqnarray}
\noindent
Due to energy denominators the positive energy part (first term in eq. (5) ) should give, and in 
fact does, the dominant contribution to the amplitude.

The Roper contribution will be included on top of the rescattering term. 
As we will see the relevance of the Roper mechanism might show
if not as a large absolute contribution yes with a large interference
with the rescattering term.
For the evaluation of the rescattering term we follow ref. \cite{11}.
We shall use the $\lambda_1$ parameter due to Hamilton 
\cite{36}, which for the half off-shell situation that we encounter here
gives a larger value than the on shell $\lambda_1^{on-shell}=0.0075$. 
In our case
\begin{eqnarray}
\lambda_1(q,p_{\pi})=-\ \frac{1}{2}\ (1+\epsilon)\ m_{\pi}
\ \left(a_{sr}+a_{\sigma}\ \frac{m_{\sigma}^2}
{m_{\sigma}^2-(q+p_{\pi})^2}\right)
\end{eqnarray}
\noindent
with $\epsilon=m_{\pi}/M$ being $M$ the nucleon mass, 
$a_{\sigma}=0.220 m_{\pi}^{-1}$, $a_{sr}=-0.233 m_{\pi}^{-1}$
and $m_{\sigma}=550 \ MeV$. Note the q here has opposite sign to the one in 
ref. \cite{11}.
    
The complex structure of the amplitude in eq. (4) makes the evaluation
of Final/Initial State Interactions Effects (FSI/ISI) a really hard task.
We will not attempt here such a calculation and will content ourselves with
the evaluation of the cross section without FSI/ISI. With these effects being
very important near threshold, we can not make strong statements about
the exact role played by the new mechanism but our hope is that we can get
at least an indication of its relevance.
%, and
% infer that probably some other, yet unexplored, mechanisms
%could be of importance for the understanding of this reaction. Do not
%forget that what we have here is a very low cross section process and that 
%non\--dominant mechanisms can be important due to interference.
%With those assumptions it is now an easy matter to evaluate 
%the cross section.

\section{Results, discussion and conclusion.} 

In the following we show and comment the theoretical results obtained with 
different sets of values for  $g_{1 \pi \pi N N^*}$ and   
$g_{2 \pi \pi N N^*}$.
 
In Table 1 we use
\mbox{$g_{1 \pi \pi N N^*}=7.27\ GeV^{-1}$} 
and \mbox{$g_{2 \pi \pi N N^*}=0$} which are the values favoured in the 
analysis of ref. \cite{32}.
As we see, the contribution of the Roper mechanism is by itself very small.
The rescattering contribution alone is also small but, as shown in
ref. \cite{11}, in this case  FSI/ISI would bring theoretical 
predictions into a fairly good
agreement with experimental data.
When one takes the Roper and the rescattering terms together  
the interference gives a reduction
of the rescattering prediction by roughly a factor of three. Thus, and
although the Roper contribution alone is too small, the net effect is,
through interference,
to reduce significantly the contribution of the dominant rescattering term.

In Table 2 we use  \mbox{$g_{1 \pi \pi N N^*}=12.7\ GeV^{-1}$} 
and \mbox{$g_{2 \pi \pi N N^*}=-1.98\ 
GeV^{-1}$}. This set of values is quoted in ref. \cite{32} as 
compatible with the experimental errors in the 
\mbox{$\pi^- p \rightarrow \pi^+ \pi^- n$} reaction. Now the contribution
of the Roper mechanism is comparable, though smaller, to the one of the 
rescattering term.
The interference between the two is destructive and the net result is a
very small cross section compared to the data.

In Table 3 the results shown correspond to 
 \mbox{$g_{1 \pi \pi N N^*}=-12.7\ GeV^{-1}$} 
and 
%\mbox{
$g_{2 \pi \pi N N^*}=1.98\ 
GeV^{-1}$
%}
. This is as before but with  opposite signs. 
Now the interference is constructive and the results at some 
energies are  bigger
than the data.

In Table 4 we have
\mbox{$g_{1 \pi \pi N N^*}=0$} 
and 
%\mbox{
$g_{2 \pi \pi N N^*}= 2.678\ 
 GeV^{-1}$
%}
. The contribution of the Roper mechanism is very small but again the interference
increases the results obtained with the rescattering term.

One can also choose a set of values for which the Roper mechanism alone
 overwhelms the data. This is done in Table 5 where we have
used
\mbox{$g_{1 \pi \pi N N^*}=-95.74\ GeV^{-1}$} 
and 
%\mbox{
$g_{2 \pi \pi N N^*}=34.61\ GeV^{-1}$
%} 
corresponding to 
one of the extremes of the ellipse. One would not expect FSI/ISI effects
to bring the results closer to experiment in 
this case  and such extreme situations have to be discarded. In  fact
these extreme cases are also excluded by the analysis of the
\mbox{$\pi^- p \rightarrow \pi^+ \pi^- n$} reaction. 

We mention once again that we are not including FSI/ISI effects 
in the calculation.
Thus, all the results presented here
have to be taken with due caution as we know these effects are very 
important.

In spite of that crude approximation, we think that from the above
 results it emerges the fact that the Roper mechanism
introduced here can  be relevant for the understanding of the 
$p p \rightarrow p p \pi^0$ reaction. Even for the cases where the 
Roper contribution alone is too small, the interference with the
dominant rescattering term is important. This situation is reminiscent 
of the role played by the Born or one\--body term considered in
calculations where FSI/ISI are included. 

The second teaching of these calculations, is that, even at the qualitative
level that we have analyzed the reaction, one can certainly exclude a wide range
of values of $g_{1 \pi \pi N N^*}$ and $g_{2 \pi \pi N N^*}$ of the ellipse of fig. 4
 allowed by the $N^*$ decay into two isoscalar s-wave pions. It is clear that
 the values situated towards the extreme of the ellipse can easily be discarded
 and only values around the origin could be compatible with experiment,
 provided other mechanisms give sizeable contributions to the reaction.
  It is rewarding
 to see that such conclusions are in agreement with findings in ref. \cite{32}
 coming from a more detailed analysis for the $\pi N \rightarrow \pi \pi N$
 reaction.
 
 A more quantitative analysis of the mechanism discussed would be advisable
 although the FSI/ISI corrections would require lengthy calculations.
 At the present time, where so many different mechanisms are suggested, most
 of them claiming an explanation of the experiment, we feel that it suffices
 to show that this mechanism is there and that its interference with other 
 mechanisms can completely change the results obtained  ignoring it.
 With our knowledge about this reaction increasing with time and different
 mechanisms settling down on a firm basis, a future detailed study taking
 all these mechanisms into consideration would be an interesting task to
 tackle.

\vspace{2cm}
Acknowledgements: This work has been partly supported by DGICYT, contract
no. PB96-0753 and Junta de Castilla y Leon contract no. SA73/98.

\newpage

\begin{table}
\caption{Cross sections for the $p p\rightarrow p p \pi^0 $ reaction
evaluated for different values of $\eta=p_{\pi\ max}/m_{\pi}$. 
Here $g_{1 \pi \pi N N^*}=7.27\ GeV^{-1} $ 
and $g_{2 \pi \pi N N^*}=0 $ (Point 1 in fig. 4). 
We show results for the Roper mechanism alone
($\sigma_{Roper}$), rescattering alone ($\sigma_{Rescat.}$), and the
full calculation($\sigma_{Total}$). For comparison we also show experimental
data taken from ref. [5] 
All cross sections are in microbarns.}

%\vspace{2.0cm}
\begin{center}
\nobreak
\begin{tabular}[t]{ |c|  c|  c|  c|  c| } 
\hline
 \large $\eta$ & \large $\sigma_{Roper}$ 
 & \large $\sigma_{Rescat.}$ &
\large $\sigma_{Total}$ & \large $\sigma_{Exp.}$
\\ 
\hline   
%\hline
%
0.203       & $1.7\ 10^{-2}$  & 0.10   & $3.7\ 10^{-2}$  & $0.70\ \pm \ 0.05$   

                                           \\   
\hline              
0.306 & $8.5\ 10^{-2}$ & 0.52  & 0.19  & $1.91\ \pm \ 0.05$  
    
                                       \\  
\hline 
0.407 & 0.26 & 1.6  & 0.56 & $3.83\ \pm \ 0.11$   
                
                               \\                 
\hline
0.517 & 0.65 & 3.9 & 1.4 & $6.18\ \pm  \ 0.20$ 
   
                                             \\                   
\hline  
\end{tabular}
\end{center}
%\label{Siegtable}
\end{table}

\begin{table}
\caption{Same as Table 1 but with
 $g_{1 \pi \pi N N^*}=12.7\ GeV^{-1} $ and 
 $g_{2 \pi \pi N N^*}=-1.98\ 
GeV^{-1}$} (Point 2 in fig. 4).

%\vspace{2.0cm}
\begin{center}
\nobreak
\begin{tabular}[t]{ |c|  c|  c|  c|  c| } 
\hline
 \large $\eta$ & \large $\sigma_{Roper}$ 
 & \large $\sigma_{Rescat.}$ &
\large $\sigma_{Total}$ & \large $\sigma_{Exp.}$
\\ 
\hline   
%\hline
%
0.203       & $6.6\ 10^{-2}$  & 0.10   & $4.2\ 10^{-3}$  & $0.70\ \pm \ 0.05$   

                                           \\   
\hline              
0.306 & 0.33 & 0.52  & $2.1\ 10^{-2}$  & $1.91\ \pm \ 0.05$  
    
                                       \\  
\hline 
0.407 & 1.0 & 1.6  & $6.1\ 10^{-2}$ & $3.83\ \pm \ 0.11$   
                
                               \\                 
\hline
0.517 & 2.6 & 3.9 & 0.15 & $6.18\ \pm  \ 0.20$ 
   
                                             \\                   
\hline  
\end{tabular}
\end{center}
%\label{Siegtable}
\end{table}

\begin{table}
\caption{Same as Table 1 but with \mbox{$g_{1 \pi \pi N N^*}=-12.7\ GeV^{-1}$} 
and \mbox{$g_{2 \pi \pi N N^*}=1.98\ 
GeV^{-1}$}} (Point 3 in fig. 4).

%\vspace{2.0cm}
\begin{center}
\nobreak
\begin{tabular}[t]{ |c|  c|  c|  c|  c| } 
\hline
 \large $\eta$ & \large $\sigma_{Roper}$ 
 & \large $\sigma_{Rescat.}$ &
\large $\sigma_{Total}$ & \large $\sigma_{Exp.}$
\\ 
\hline   
%\hline
%
0.203       & $6.6\ 10^{-2}$  & 0.10   & 0.33  & $0.70\ \pm \ 0.05$   

                                           \\   
\hline              
0.306 & 0.33 & 0.52  & 1.7  & $1.91\ \pm \ 0.05$  
    
                                       \\  
\hline 
0.407 & 1.0 & 1.6  & 5.1 & $3.83\ \pm \ 0.11$   
                
                               \\                 
\hline
0.517 & 2.6 & 3.9 & 12.8 & $6.18\ \pm  \ 0.20$ 
   
                                             \\                   
\hline  
\end{tabular}
\end{center}
%\label{Siegtable}
\end{table}

\begin{table}
\caption{Same as Table 1 but with $g_{1 \pi \pi N N^*}=0$ and 
$g_{2 \pi \pi N N^*}=2.678\ 
GeV^{-1}$} (Point 4 in fig. 4).

%\vspace{2.0cm}
\begin{center}
\nobreak
\begin{tabular}[t]{| c|  c|  c|  c|  c| } 
\hline
 \large $\eta$ & \large $\sigma_{Roper}$ 
 & \large $\sigma_{Rescat.}$ &
\large $\sigma_{Total}$ & \large $\sigma_{Exp.}$
\\ 
\hline   
%\hline
%
0.203       & $1.8\ 10^{-3}$  & 0.10   & 0.13 & $0.70\ \pm \ 0.05$   

                                           \\   
\hline              
0.306 & $9.1\ 10^{-3}$ & 0.52  & 0.67  & $1.91\ \pm \ 0.05$  
    
                                       \\  
\hline 
0.407 & $2.8\ 10^{-2}$ & 1.6  & 2.0 & $3.83\ \pm \ 0.11$   
                
                               \\                 
\hline
0.517 & $7.3\ 10^{-2}$ & 3.9 & 5.0 & $6.18\ \pm  \ 0.20$ 
   
                                             \\                   
\hline  
\end{tabular}
\end{center}
%\label{Siegtable}
\end{table}

\begin{table}
\caption{Same as Table 1 but with $g_{1 \pi \pi N N^*}=-95.74\ GeV^{-1}$ and 
$g_{2 \pi \pi N N^*}=34.61\ 
GeV^{-1}$} (Point 5 in fig. 4).

%\vspace{2.0cm}
\begin{center}
\nobreak
\begin{tabular}[t]{| c|  c|  c|  c|  c| } 
\hline
 \large $\eta$ & \large $\sigma_{Roper}$ 
 & \large $\sigma_{Rescat.}$ &
\large $\sigma_{Total}$ & \large $\sigma_{Exp.}$
\\ 
\hline   
%\hline
%
0.203       & 5.0  & 0.10   & 6.5 & $0.70\ \pm \ 0.05$   

                                           \\   
\hline              
0.306 & 26 & 0.52  & 33  & $1.91\ \pm \ 0.05$  
    
                                       \\  
\hline 
0.407 & 78 & 1.6  & 102 & $3.83\ \pm \ 0.11$   
                
                               \\                 
\hline
0.517 & 197 & 3.9 & 256 & $6.18\ \pm  \ 0.20$ 
   
                                             \\                   
\hline  
\end{tabular}
\end{center}
%\label{Siegtable}
\end{table}

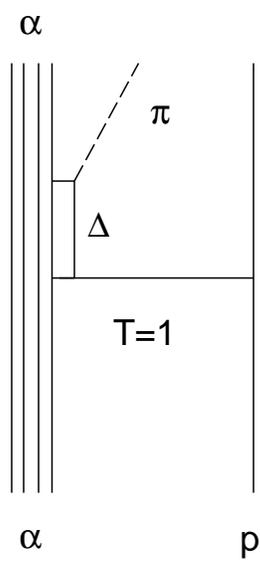
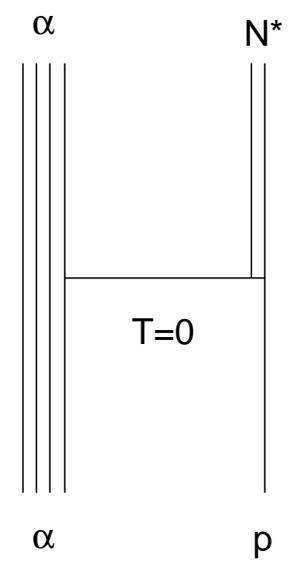
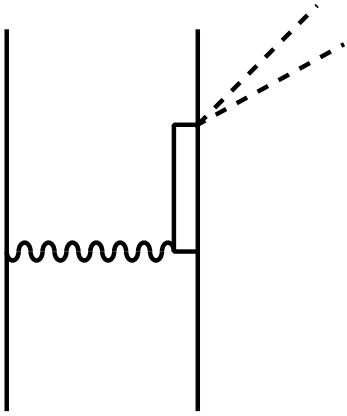
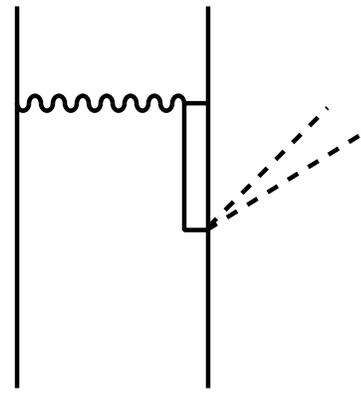
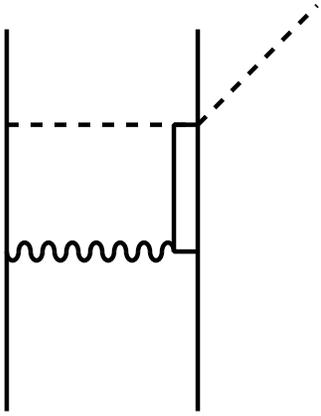
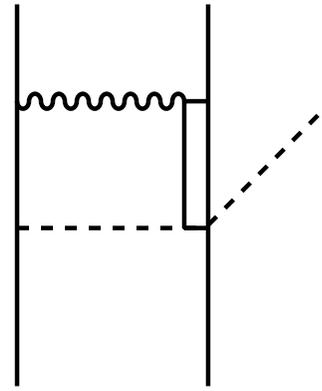
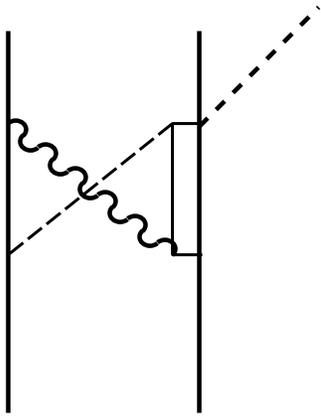
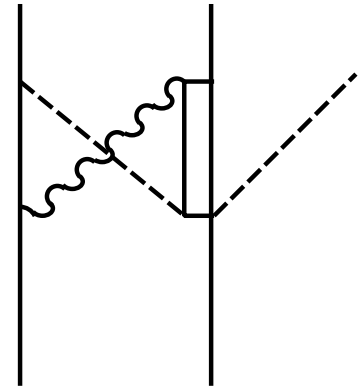
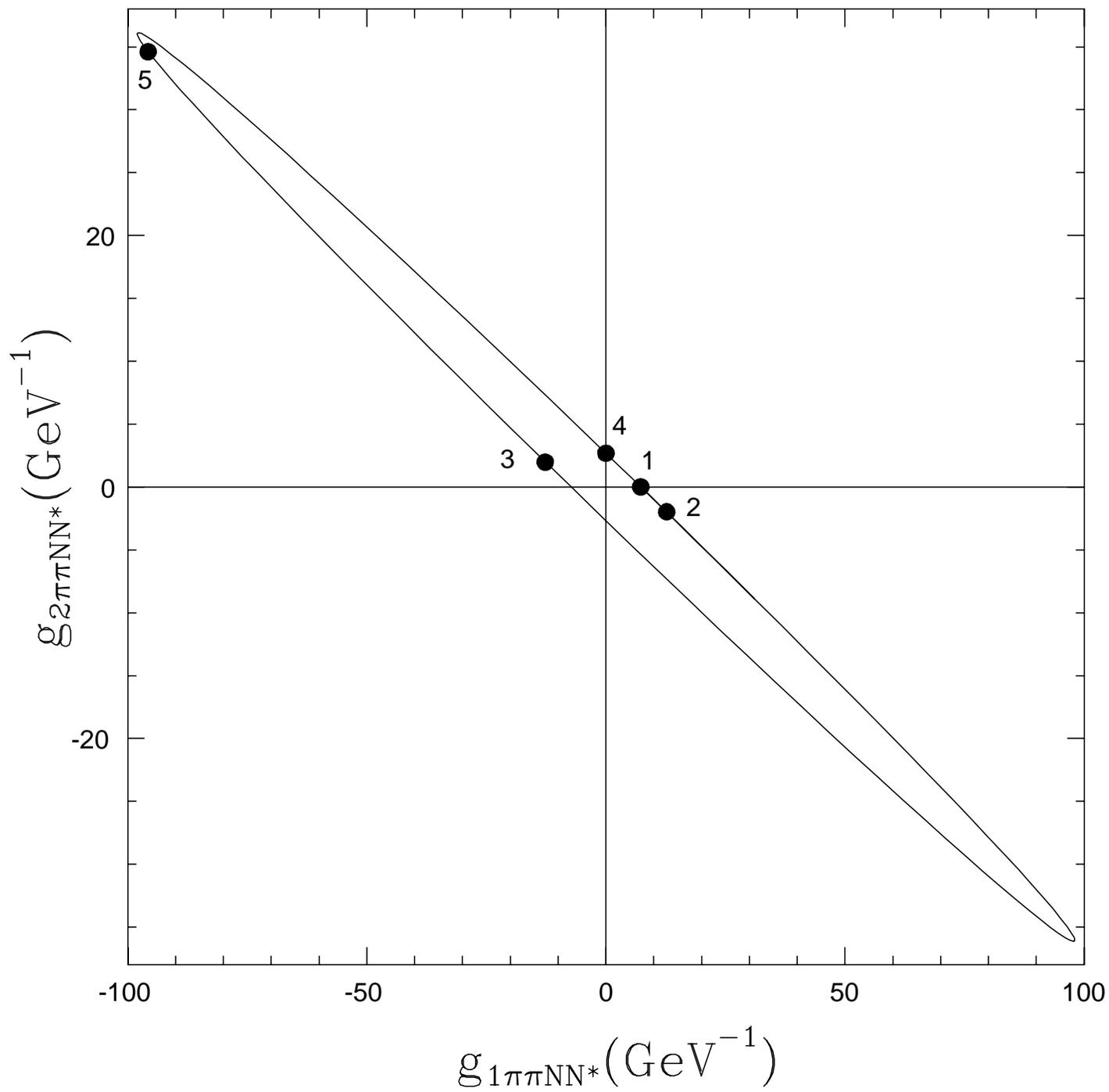
\end{document}